\documentstyle[sprocl]{article}

\def\Journal#1#2#3#4{{#1} {\bf #2}, #3 (#4)}

\def\PLB{{\it Phys. Lett.}  B}

\def\PRD{{\it Phys. Rev.} D}

\def\be{\begin{equation}}
\def\ee{\end{equation}}
\def\bea{\begin{eqnarray}}
\def\eea{\end{eqnarray}}

\begin{document}
\title{RELATIVISTIC GRAVITY WITH A DYNAMICAL PREFERRED FRAME}
\author{D. MATTINGLY AND T. JACOBSON}
\address{Physics Department, University of Maryland,
College Park,\\ MD 20742-4111, USA\\
E-mail: davemm@physics.umd.edu, jacobson@physics.umd.edu}
\maketitle
{\em Talk presented by D. Mattingly at CPT01; the Second Meeting
on CPT and Lorentz Symmetry, Bloomington, Indiana, 15-18 Aug.
2001.}

\section{Introduction}
While general relativity possesses local Lorentz invariance, both
canonical quantum gravity \cite{GPDispersion} and string theory
\cite{KSstring} suggest that Lorentz invariance may be broken at
high energies. Broken Lorentz invariance has also been postulated
as an explanation for astrophysical anomalies such as the missing
GZK cutoff \cite{CGGZK}. Therefore, we seek an effective field
theory description of gravity where Lorentz invariance is broken.
We will construct a candidate theory and then briefly discuss some
of the implications.

\section{Construction of the Effective Field Theory}
Since we have observational evidence only in a small range of
energies relative to the expected Planck energy scale of quantum
gravity, it is plausible that boost invariance is broken and as
yet unobserved. Rotation invariance has been uniformly explored
however, and hence for now we assume rotation symmetry is
preserved. A structure that preserves rotation invariance and not
boost invariance is that of a preferred frame or ``aether" which
is mathematically realized by a unit future timelike vector field
$u^a$.

There are many possible Lorentz breaking effects involving $u^a$.
For example, a matter field might possess a modified dispersion
relation $\omega^2 =
|\vec{k}|^2-k_0^{-2}|\vec{k}|^4$, where
$k_0$ is a constant (mostly likely of order the Planck energy)
that sets the scale of the Lorentz breaking.  A Lagrangian that
gives the above dispersion for a scalar field $\phi$ is
\begin{equation}\label{lagrangian}
L_{\phi}=\frac {1} {2} ( \nabla^a \phi \nabla_a \phi + k_0^{-2}
(D^2 \phi)^2)
 \end{equation}
where the spatial Laplacian $D^2$ is defined by
\begin{equation}\label{spatiallaplacian}
D^2 \phi = -D^a D_a \phi=-q^{ac} \nabla_c(q_a^b\nabla_b\phi)
 \end{equation}
and the spatial metric $q_{ab}$ is defined by
\begin{equation}\label{spatialprojector}
 q_{ab}=-g_{ab}+u_au_b.
 \end{equation}

This type of construction suffices for a Lorentz breaking theory
in flat spacetime. However, if we try to couple to gravity by
adding this action to the Einstein-Hilbert action, the resulting
theory is inconsistent because the fixed vector $u^a$ introduces
prior geometry and therefore violates general covariance. (The
same would be true for a Lorentz breaking tensor field.) If
general covariance is violated in this way then the stress tensor
for the matter field will not be conserved, rendering the theory
unviable. In order to preserve general covariance it is necessary
that $u^a$ become dynamical.

Since we have no underlying theory that tells us what form the
$u^a$ kinetic terms might take, we follow the spirit of effective
field theory and make a derivative expansion for the $u^a$
Lagrangian. Including all terms with up to two derivatives of
$u^a$ and $g_{ab}$, and neglecting total divergences, we have
\begin{eqnarray}\label{ulagrangian}
  L_{g,u} & = & a_0 - a_1R -a_2R_{ab}u^au^b \hfill \\
  & & -b_1F_{ab}F^{ab}-b_2(\nabla_au_b)(\nabla^au^b) -
  b_3\dot{u}^a\dot{u}_a+\lambda(g_{ab}u^au^b-1)
\end{eqnarray}
where $\dot{u}^a=u^b\nabla_bu^a$ and $F_{ab}=2\nabla_{[a}u_{b]}$.
$\lambda$ is a Lagrange multiplier such that the unit constraint
on $u^a$ is enforced dynamically as an equation of motion. The
theory with Lagrangian density $\sqrt{-g}(L_u + L_{\phi})$
is generally covariant since it involves no fixed
background structures. Such vector-tensor theories have been
previously studied by several authors, both with
normalized \cite{gasperini,KSeffective,jmaether} and
non-normalized \cite{Will,ClayMoff} $u^a$.
The dynamical aether field has gravitational consequences in
addition to the non-gravitational effects of the matter-aether
coupling. We turn now to a brief discussion of some of these
consequences.

\section{Observational Consequences}

\subsection{Field of Static Bodies}
The static spherically symmetric solutions for the gravitational
field of a body such as the sun are modified by the introduction
of $u^a$.  Solar system tests will therefore place constraints on
the coefficients in the theory.  Unfortunately, neither the
general spherically symmetric static solution nor the PPN
parameters for this theory are currently known.

The static spherically symmetric solutions for the case where only
$a_1,b_1\neq0$ have all been found, however \cite{jmaether}.  In
this case, the theory is equivalent to a sector of
Einstein-Maxwell charged dust theory where the dust has a charge
to mass ratio $-1/2\sqrt{b_1}$. There exists a black hole solution
of the Reissner-Nordstrom form where the electric charge Q is
replaced by the charge of the aether dust that fell into the black
hole.  This raises the possibility that the general theory may
also introduce a new one parameter family of black hole solutions.

\subsection{E\"{o}tv\"{o}s Experiments}
The coupling of matter to $u^a$ can result non-geodesic free-fall
trajectories for particles.
This violation of the equivalence principle
should be detectable in principle by E\"{o}tv\"{o}s experiments.
However, if $k_0$ is of order the Planck scale current experiments
are not sensitive enough to detect this violation if it arises
from couplings like those in (\ref{lagrangian}).

\subsection{Gravitational Waves}
The metric has the usual transverse traceless (TT) modes where the
aether is unperturbed, however their speed is generically
modified. In addition there are generically three aether-gravity
modes. For example, the table below lists the speeds, relevant
metric polarization components $\epsilon_{\mu\nu}$, and relevant
aether polarization components $w^\alpha$, for the theory with
only $a_1,b_2\neq 0$ in Lorentz gauge. The wave vector is of the
form $(k_0,0,0,k_3)$, $v$ is the wave speed $k_0/k_3$, $I,J$ run
from 1 to 2, and $\tau=b_2/a_1$.

\begin{table} [ht]
  \centering
  \label{polarizations}
\begin{tabular}{|llll|}
    \hline
\rule[-3mm]{0mm}{7mm}
Mode & $v^2$& Metric Components & Aether Components\\
    \hline
    \rule[-3mm]{0mm}{7mm}
Transverse Traceless & $\frac {1} {1+\tau}$ & $\epsilon_{IJ}$ & -\\
    \rule[-3mm]{0mm}{7mm}
Transverse Vector & $\frac {2+\tau} {2+2\tau}$ & $\epsilon_{0I},\epsilon_{3I} $ & $w^I$\\
    \rule[-3mm]{0mm}{7mm}
Longitudinal & $\frac {2+\tau-\tau^2} {2+\tau}$ &
 $\epsilon_{00},\epsilon_{03},\epsilon_{33},\epsilon_{IJ}$ & $w^0,w^3$\\
    \hline
\end{tabular}
\end{table}

Astrophysical sources, such as coalescing black holes or neutron
stars, may couple only weakly to the new modes.  Even if so, a
gravitational wave observatory could potentially still detect the
TT modes travelling at a speed other than c by comparing time of
arrival data with non-gravitational signals from the same event.

\subsection{Cosmology}
With a consistent gravity-aether-matter theory, one can look at
the cosmological implications of Lorentz symmetry breaking.  There
are effects due to both the aether stress tensor and the modified
field equations for matter fields.  Assuming the aether frame
coincides with the isotropic frame of a Robertson-Walker metric,
the aether stress tensor has at most two two terms.  The first
term is proportional to the Einstein tensor and hence renormalizes
$G$. The second term, which is non-vanishing only if there is
spatial curvature, is that of a perfect fluid with pressure equal
to $-1/3$ times the energy density (like the spatial curvature
term in the Friedmann equations).  The aether itself therefore
affects the cosmological expansion rate.

For a Lorentz-violating matter coupling like that in
(\ref{lagrangian}) the equation of state is different at high
energies.  This does not lead to dramatic consequences for
cosmological evolution at energies up to $k_0$
\cite{jmcosmology}.  However in an inflationary scenario where
cosmological scales were once much smaller than $k_0$, modified
dispersion could have important effects on the power spectrum of
the primordial metric fluctuations (c.f. ref.
\cite{llcosmology} and references therein).  Results for energies
above $k_0$ should be treated with caution though as the aether
theory is a low energy effective theory and is most likely not
applicable for energies at or above $k_0$.

\section{Viability Issues}
\subsection{Stability}
If the aether theory is to be viable, it must be energetically
stable.  In a generally covariant theory the energy is given by
the boundary term in the Hamiltonian; this is the ADM mass in
general relativity.  The energy for the aether theory with
Lagrangian (\ref{ulagrangian}) has not yet been calculated.  It
may not be the ADM mass, due to the aether ``kinetic term'' which
introduces additional $(\partial g)^2$ terms via the connection
components in $\nabla_au_b$.  These terms persist even at spatial
infinity because $u^a$ is a unit vector.  The usual positive
energy theorem may therefore no longer apply (as it assumes that
the energy is the ADM mass).  What we need is a new positive
energy theorem, using the correct definition of energy, to
establish the range of parameters in the Lagrangian
(\ref{ulagrangian}) for which the energy is always positive.  It
may be generally only positive in the asymptotic aether frame,
which could be enough to guarantee stability of the theory.

\subsection{Shocks}
In the theory with only $a_1,b_1\neq0$, generic initial data will
develop shocks where the integral curves of $u^a$ cross,
signaling a breakdown of the effective theory
\cite{jmaether,clayton}. The underlying source of these shocks is
the insensitivity of the action to the symmetric derivative of
$u^a$. It is hoped that including terms involving
the symmetric derivative will cure this
problem.

\section{Conclusion}
The introduction of Lorentz breaking into general relativity
demands the addition of another dynamical field. Since this extra
field has its own stress tensor and thereby gravitational
consequences, the Lorentz breaking effects for matter are only
half the story. Any new theory must be energetically stable and
not contradict observations. The aether theory of a preferred
frame discussed here is a candidate for such a theory, but much
work remains to be done.\\

\noindent {\it Work supported in part by the NSF
under grant PHY98-00967.}

\section*{References}


\begin{thebibliography}{99}
\bibitem{GPDispersion}
J. ~Alfaro, H. ~Morales-Tecotl, and L. ~Urrutia, ``Loop quantum
gravity and light propagation'', [hep-th/0108061].

\bibitem{KSstring}
V.~A. ~Kostelecky and S. ~Samuel, ``Spontaneous breaking of
Lorentz symmetry in string theory'',  \Journal{\PRD} {39} {683}
{1989}.

\bibitem{CGGZK}
S. ~Coleman and S. ~Glashow, ``High-energy tests of Lorentz 
invariance," \Journal{\PRD} {59} {116008} {1999}.

\bibitem{gasperini}
M. ~Gasperini, ``Singularity prevention and broken Lorentz
symmetry'', \Journal{Class. Quantum Grav.} {4} {485} {1987}.

\bibitem{KSeffective}
V.~A.~Kostelecky and S.~Samuel, ``Gravitational Phenomenology In
Higher Dimensional Theories And Strings,'' \Journal{\PRD} {40}
{1886} {1989}.

\bibitem{jmaether}
T. ~Jacobson and D. ~Mattingly, ``Gravity with a dynamical
preferred frame'', \Journal{\PRD} {64} {024028} {2001}.

\bibitem{Will}
C. ~Will, \textit{Theory and experiment in gravitational physics},
Cambridge University Press, 1993.

\bibitem{ClayMoff}
M.~A.~Clayton and J.~W.~Moffat, ``Dynamical Mechanism for Varying
Light Velocity as a Solution to Cosmological Problems,''
\Journal{\PLB} {460} {263} {1999}.

\bibitem{jmcosmology}
T. ~Jacobson and D. ~Mattingly, ``Generally covariant model of a
scalar field with high frequency dispersion and the cosmological
horizon problem", \Journal{\PRD} {63} {041502} {2001}.

\bibitem{llcosmology}
M. ~Lemoine, et. al., ``The stress-energy tensor for
trans-Planckian cosmology'', [hep-th/0109128].

\bibitem{clayton}
M. ~Clayton, ``Causality, shocks and instabilities in vector field
models of Lorentz symmetry breaking'', [gr-qc/0104103].






\end{thebibliography}
\end{document}